# "Cosmic Alternator" for the Onset of Large-Scale Magnetic Fields

B. Coppi and B. Basu

Massachusetts Institute of Technology, Cambridge, MA 02139 USA

**Abstract**

Magnetic field configurations extending over macroscopic scale distances are shown to be generated in rarefied collisionless plasmas when non-thermal and spatially inhomogeneous electron distributions in phase space emerge. The analyzed representative case is where, in the presence of a significant spatial gradient of the electron pressure and of a sheared magnetic field configuration, an alternating magnetic field reconnection process can develop associated with the anisotropy of the fluctuating electron temperature. The relevant process is intrinsically different from that known as the "Biermann Battery" which does not involve magnetic reconnection and requires that the unperturbed spatial gradient of the (isotropic) electron temperature be misaligned relative to that of the density gradient.

Corresponding author: B. Coppi (coppi@mit.edu)

# 1. Introduction

Magnetic reconnection processes have a key role to play in the development of important observed phenomena and events both in the laboratory and in space. The nature of these processes is radically different in the different physical regimes that are involved. The analysis that will follow addresses a wide category of weakly collisional or collisionless (under appropriate conditions) regimes where the electron thermal conductivity along the main component of the magnetic field is large and much larger than the transverse thermal conductivity. Magnetic reconnection has been commonly recognized as a process by which magnetic energy is converted into particle kinetic energy [1] but, as shown in the following analysis, the opposite kind of conversion can take place and the electron thermal energy reservoir is an important factor for the kind [2] of magnetic reconnection that can take place.

When the issue of identifying one of the possible origins of magnetic fields over large scale distances in the Universe is addressed, the process [3] represented by the "cosmic alternator", that is the subject of the following analysis, can be considered as one of the key components of a relevant theory. For this, as indicated in Section 9, an additional kind of process has to be introduced that leads to the formation of a "seed" magnetic field configuration from which a growing reconnected field configuration can emerge.

The present paper is organized as follows. In Section 2, the simplest sheared magnetic field configuration is introduced and the conditions, under which an electron population with an anisotropic temperature [4] is confined, are given. In Section 3, the limits of validity of the undertaken analysis are identified considering regimes where the longitudinal (to the magnetic field) electron thermal conductivity is relatively large. In Section 4, the linearized electron momentum conservation equation is introduced that is centered on the relevant electron pressure tensor and temperature anisotropy. A reconnecting component of the magnetic field is shown to emerge for a characteristic frequency of the considered mode that depends on the electron pressure gradient and involves anisotropic electron temperature fluctuations. In Section 5, an electron thermal energy balance equation is introduced that features a relatively large longitudinal electron thermal conductivity and removes a characteristic spatial singularity of electron temperature fluctuations. The solution of this equation is given that relates the longitudinal electron temperature fluctuation to the reconnecting magnetic field component. In Section 6, the lowest order terms of the total momentum conservation equation are identified that are relevant to the reconnection theory under consideration. In Section 7, the asymptotic matching condition of the solution, considering the longitudinal electron temperature fluctuation, in the inner region where the transverse electron thermal conductivity is important to the outer (ideal MHD) region is derived. This condition is shown to depend strongly on the gradients of the electron temperature anisotropy and leads to identify two kinds of mode. One of these is found to acquire a growth rate, as shown in Section 8, when a finite magnetic field diffusion (resistive) coefficient is introduced. In this case, the plasma current density gradient acquires a key role. In Section 9, final remarks are made concerning the need of a complementary theory for collisionless regimes that include mode-particle resonant interactions. In the same section a theoretical model is discussed that can provide an explanation for the emergence of large-scale magnetic fields in space based on a (cosmic alternator) process of the kind considered in this paper.

The relevance of particle transport to the presented reconnection theory that depends on electron temperature anisotropy is demonstrated in the Appendix.

**2. Unperturbed State**

The main points of the undertaken analysis can be brought forward by considering an unperturbed sheared magnetic field configuration that is one dimensional and a plasma sheet where only the electron pressure is anisotropic. That is, we take $\mathbf{B} = B_z(x)\mathbf{e}_z + B_y(x)\mathbf{e}_y$, with $B_y^2 \ll B_z^2 \simeq B^2$, $\mathbf{P}_e = (p_{e\|} - p_{e\perp})\mathbf{BB}/B^2 + p_{e\perp}\mathbf{I}$, $p_{e\|} \sim p_{e\perp}$, $\mathbf{P}_i = p_i\mathbf{I}$ and $p_i \sim p_{e\perp}$. In this model the plasma density $n(x)$ and the current density $J_z = (c/4\pi)(dB_y/dx)$ are peaked at $x = 0$, and

$$\frac{dp_\perp}{dx} = \frac{1}{c}(J_y B_z - J_z B_y) = 0 \tag{1}$$

at $x = 0$ as $B_y \simeq (dB_y/dx)x$, $J_y \simeq (dJ_y/dx)x$ near $x = 0$. Referring to a surface $x = x_0 \neq 0$, we consider $p_\perp \simeq p_{e\perp} + p_i \sim B_y^2$.

For the momentum conservation of the ion population,

$$\frac{dp_{i\perp}}{dx} = en\left[E_x + \frac{1}{c}(u_{iy}B_z - u_{iz}B_y)\right], \tag{2}$$

we consider two options. One where this population is electrostatically confined, that is $E_x = (dp_{i\perp}/dx)/(en)$, and the other where it is magnetically confined, that is

$$\frac{dp_{i\perp}}{dx} = en\frac{1}{c}u_{iy}B_z. \tag{3}$$

Then the electron momentum conservation equation is

$$\frac{dp_{e\perp}}{dx} = -en\left[E_x + \frac{1}{c}(u_{ey}B_z - u_{ez}B_y)\right] \tag{4}$$

and gives

$$u_{ey} = u_{ez}\frac{B_y}{B_z} - \frac{c}{B_z}\left(E_x + \frac{1}{en}\frac{dp_{e\perp}}{dx}\right). \tag{5}$$

We observe that the oscillation frequency of the modes which will be introduced will depend on the terms that have to be retained in Eq. (5) to represent the chosen unperturbed state. In view of the following analysis we conjecture the macroscopic scale distance $r_p$ defined as $1/r_p \equiv -(dp_{e\|}/dx)/p_{e\|}$ to be of the same order as the width of the plasma sheet and assume that $\beta_e \equiv 4\pi p_{e\|}/B^2$ is small both for the sake of simplicity as well as considering that it is realistic for significant magnetically confined plasmas.

## 3. Analysis Limits

The linearized analysis that is carried out involves small perturbations represented by $\hat{B}_y(x,y,z,t)$ such that $|\hat{B}_y| < |B_y|$. The reconnecting field component $\hat{B}_x$ generated over a distance $\delta_T$ that will be defined later produces a magnetic island whose width is of the order of

$$\delta_I = \left|\frac{\hat{B}_x}{(dB_y/dx)k_y}\right|^{1/2} \tag{6}$$

as we consider modes represented by

$$\hat{B}_y = \tilde{B}_y(x)\exp\left[-i(\omega t - k_y y - k_z z)\right], \tag{7}$$

where $\mathbf{k}\cdot\mathbf{B} = k_y B_y + k_z B_z = 0$ for $x = x_0$. Then we require, for the validity of the linearized approximation, that

$$|\tilde{B}_x| < \left|\frac{dB_y}{dx}k_y\right|\delta_T^2 \tag{8}$$

and, since $\partial\tilde{B}_x/\partial x \simeq -ik_y\tilde{B}_y$, Eq. (8) implies that (see Section 7)

$$\left|\frac{\tilde{B}_y}{B_y}\right| < \varepsilon_T \left|\frac{1}{B_y}\frac{dB_y}{dx}\right|\delta_T, \tag{9}$$

where $\varepsilon_T \ll 1$, and this is the most stringent condition on $|\tilde{B}_y / B_y|$.

We note that

$$\nabla \cdot \mathbf{P}_e = \mathbf{B} \cdot \nabla \left[ \frac{\mathbf{B}}{B^2} (p_{e\parallel} - p_{e\perp}) \right] + \nabla p_{e\perp} \tag{10}$$

and that the perturbed form of this is approximately,

$$\nabla \cdot \hat{\mathbf{P}}_e \simeq \hat{B}_x \frac{d}{dx} \left[ \frac{\mathbf{B}}{B^2} \Delta p_e \right] + (\mathbf{B} \cdot \nabla) \left[ \frac{\hat{\mathbf{B}}}{B^2} \Delta p_e + \frac{\mathbf{B}}{B^2} \Delta \hat{p}_e \right] + \nabla \hat{p}_{e\perp}, \tag{11}$$

where $\Delta p_e \equiv p_{e\parallel} - p_{e\perp}$.

Before going through a complete analysis, we may make the following general arguments, given that we consider weakly collisional regimes,

1) the transverse electron temperature fluctuations can be expected to be smaller than the longitudinal temperature fluctuations on account of the fact that the electron cyclotron frequency is much larger than the relevant collision frequency.

2) the longitudinal thermal conductivity can be assumed to be large and lead to the condition

$$\mathbf{B} \cdot \nabla T_{e\parallel} = 0 \tag{12}$$

that, in the perturbed form, is

$$\mathbf{B} \cdot \nabla \hat{T}_{e\parallel} + \hat{B}_x \frac{dT_{e\parallel}}{dx} = 0. \tag{13}$$

Therefore, if $\tilde{B}_x(x = x_0) \neq 0$, $\tilde{T}_{e\parallel}$ would be singular at $x = x_0$ and an operator involving a singular perturbation would need to be included in the theory in order to make $\tilde{T}_{e\parallel}(x)$ regular at $x = x_0$.

3) a dissipative transport process has to be introduced in order to obtain a growth rate.

**4. Electron Momentum Conservation Equation**

The adopted perturbed electron momentum conservation equation is

$$-e\left[\hat{\mathbf{E}} + \frac{1}{c}\left(\hat{\mathbf{u}}_e \times \mathbf{B} + \mathbf{u}_e \times \hat{\mathbf{B}}\right)\right] - \frac{1}{n}\nabla \cdot \hat{\mathbf{P}}_e + \left(\frac{\nabla p_{e\perp}}{n}\right)\frac{\hat{n}}{n} = 0 \tag{14}$$

and is used to evaluate the reconnecting component $\hat{B}_x$ from

$$i\frac{\omega}{c}\hat{B}_x = \left(\nabla \times \hat{\mathbf{E}}\right)_x. \tag{15}$$

Then

$$\left(\nabla \times \hat{\mathbf{E}}\right)_x \simeq -\frac{1}{c}\left(\mathbf{B} \cdot \nabla \hat{u}_{ex} - \mathbf{u}_e \cdot \nabla \hat{B}_x\right) - \frac{1}{en}\left\{\nabla \times \left(\nabla \cdot \hat{\mathbf{P}}_e\right)\right\}_x, \tag{16}$$

where

$$\mathbf{u}_e \simeq \left\{-\frac{c}{B}\left(E_x + \frac{1}{en}\frac{dp_{e\perp}}{dx}\right) + u_{ez}\frac{B_y}{B_z}\right\}\mathbf{e}_y + u_{ez}\mathbf{e}_z, \tag{17}$$

as $k_y/k_z = -B_z/B_y$ for $x = x_0$. The last term in Eq. (16) becomes

$$-\frac{1}{ne}\left\{ik_y\left[\left(\frac{d}{dx}\Delta p_e\right)\frac{\hat{B}_x}{B} + \frac{1}{B}\mathbf{B}\cdot\nabla\left(\Delta\hat{p}_e\right)\right] - ik_z\left[\hat{B}_x\frac{d}{dx}\left(\Delta p_e\frac{B_y}{B^2}\right) + \frac{B_y}{B^2}\mathbf{B}\cdot\nabla\left(\Delta\hat{p}_e\right)\right]\right\}, \tag{18}$$

and, as a result,

$$i\frac{\omega}{c}\hat{B}_x \simeq -\frac{i}{c}\left[(\mathbf{k}\cdot\mathbf{B})\hat{u}_{ex} + k_y\frac{c}{B}\left(E_x + \frac{1}{en}\frac{dp_{e\|}}{dx}\right)\hat{B}_x\right] + \frac{1}{enB}k_y(\mathbf{k}\cdot\mathbf{B})\left(\hat{p}_{e\|} - \hat{p}_{e\perp}\right) \tag{19}$$

and

$$i\left[\omega + k_y\frac{c}{B}\left(E_x + \frac{1}{en}\frac{dp_{e\|}}{dx}\right)\right]\hat{B}_x \simeq -i(\mathbf{k}\cdot\mathbf{B})\left[\hat{u}_{ex} + ik_y\frac{c}{enB}\left(\hat{p}_{e\|} - \hat{p}_{e\perp}\right)\right]. \tag{20}$$

Therefore, we consider

$$\omega = \omega_0 = -k_y\frac{c}{B}\left(E_x + \frac{1}{en}\frac{dp_{e\|}}{dx}\right) \tag{21}$$

and in the case when the ions are electrostatically confined,

$$\omega_0 = -k_y \frac{c}{enB} \frac{d}{dx}(p_{i\perp} + p_{e\|}).$$

Then we take

$$\hat{u}_{ex} \simeq -ik_y D_B^e \frac{1}{nT_{e\|}} (\hat{p}_{e\|} - \hat{p}_{e\perp}), \tag{22}$$

where $D_B^e \equiv cT_{e\|}/(eB)$ and it is clear that the anisotropy of the perturbed electron pressure is the factor that allows modes with $\omega = \omega_0$ to exist.

Clearly

$$\tilde{p}_{e\|} - \tilde{p}_{e\perp} = \tilde{n}(T_{e\|} - T_{e\perp}) + n(\tilde{T}_{e\|} - \tilde{T}_{e\perp}) \equiv \tilde{n}(\Delta T)_e + n(\Delta \tilde{T}_e) \tag{23}$$

and we shall analyze, at first, the case where $|\tilde{T}_{e\perp}| \ll |\tilde{T}_{e\|}|$ as we consider modes involving perturbed electron populations for which $\mu_B = m_e \mathrm{v}_y^2/B$ is an adiabatic invariant.

Therefore, if $(\Delta T)_e \equiv T_{e\|} - T_{e\perp} \neq 0$, $\tilde{p}_{e\|} - \tilde{p}_{e\perp} \simeq n\tilde{T}_{e\|}$ and

$$\tilde{u}_{ex} \simeq -ik_y D_B^e \left\{ \frac{1}{T_{e\|}} (\Delta T)_e \frac{\tilde{n}}{n} + \frac{\tilde{T}_{e\|}}{T_{e\|}} \right\}. \tag{24}$$

Moreover,

$$\tilde{n}_e \simeq -i \frac{\tilde{u}_{ex}}{\omega_0 - \mathbf{k} \cdot \mathbf{u}_e} \frac{dn}{dx} \tag{25}$$

in the case where particle transport across the field can be neglected. Then, referring to Eq. (24)

$$\tilde{u}_{ex} \simeq -ik_y D_B^e \left\{ \frac{1}{T_{e\|}} (\Delta T)_e \left( -\tilde{\xi}_x \frac{1}{n} \frac{dn}{dx} \right) + \frac{\tilde{T}_{e\|}}{T_{e\|}} \right\}. \tag{26}$$

where

$$-i(\omega_0 - \mathbf{k} \cdot \mathbf{u}_e)\tilde{\xi}_x = \tilde{u}_{ex}. \tag{27}$$

We note also that

$$\bar{\omega}_e \equiv \omega_0 - \mathbf{k} \cdot \mathbf{u}_e \simeq -k_y D_B^e \frac{1}{nT_{e\|}} \frac{d}{dx}\left[n(\Delta T)_e\right]. \tag{28}$$

Therefore, Eq. (26) reduces to

$$ik_y D_B^e \frac{1}{nT_{e\|}} \frac{d}{dx}\left[n(\Delta T)_e\right]\tilde{\xi}_x \simeq ik_y D_B^e \left\{\frac{1}{T_{e\|}}(\Delta T)_e \left(\tilde{\xi}_x \frac{1}{n}\frac{dn}{dx}\right) - \frac{\tilde{T}_{e\|}}{T_{e\|}}\right\} \tag{29}$$

and gives

$$\left[\frac{d}{dx}(\Delta T)_e \tilde{\xi}_x\right] \simeq -\tilde{T}_{e\|} \tag{30}$$

showing the importance of $d(\Delta T)_e / dx$ in the relationship between $\tilde{T}_{e\|}$ and $\tilde{\xi}_x$.

## 5. Electron Thermal Energy Transport

We note that Eq. (13) can be considered to be valid for $|x - x_0| \gg \delta_T$ giving

$$\tilde{T}_{e\|} \simeq i \frac{\tilde{B}_x}{(dB_y/dx)} \frac{1}{k_y \delta_T} \left(\frac{dT_{e\|}}{dx}\right) \frac{1}{\bar{x}}, \tag{31}$$

where $\delta_T$ represents the width of the layer where the $\bar{x} = 0$ singularity appearing in Eq. (31) is removed and $\bar{x} \equiv (x - x_0)/\delta_T$. The reconnecting field solution $\tilde{B}_x$ that we analyze is nearly constant within the region $\delta_T$, extends outside it and can be represented as

$$\tilde{B}_x \simeq \tilde{B}_x^0 \left[1 + \varepsilon_T \varphi_0(\bar{x})\right], \tag{32}$$

where $\varepsilon_T \ll 1$.

A simple thermal energy transport equation, that has the required characteristics to remove the indicated singularity is the following

$$D_\perp^e \frac{d^2 \tilde{T}_{e\|}}{dx^2} - D_\|^e k_\| \left(k_\| \tilde{T}_{e\|} - i \frac{dT_{e\|}}{dx} \frac{\tilde{B}_x}{B}\right) \simeq 0, \tag{33}$$

where $k_\parallel \simeq k_y (B'_y / B)(x - x_0)$, $B'_y \equiv (dB_y / dx)$ and we assume that $|\omega_0| < D^e_\parallel k_\parallel^2$, $D^e_\parallel$ and $D^e_\perp$ representing the longitudinal and transverse thermal conductivities, respectively. Consequently,

$$\delta_T^4 = \frac{D^e_\perp}{D^e_\parallel} \left( \frac{B}{k_y B'_y} \right)^2 \tag{34}$$

and we shall consider

$$\varepsilon_T \geq \delta_T \left| \frac{B'_y}{B_y} \right|. \tag{35}$$

Clearly, Eq. (33) can be rewritten as

$$\frac{d^2 \tilde{T}_{e\parallel}}{d\bar{x}^2} - \bar{x} \left( \bar{x} \tilde{T}_{e\parallel} - \tilde{T}_M \right) \simeq 0 \tag{36}$$

where $\tilde{T}_M \equiv i\tilde{B}_x (dT_{e\parallel} / dx) / (k_y \delta_T B'_y)$. If we define

$$\bar{Y} \equiv \frac{\tilde{T}_{e\parallel}}{\tilde{T}_M} \simeq \frac{\tilde{T}_{e\parallel}}{i\tilde{B}_x^0} \left( \frac{dB_y}{dx} / \frac{dT_{e\parallel}}{dx} \right)(k_y \delta_T), \tag{37}$$

Eq. (36) becomes

$$\frac{d^2 \bar{Y}}{d\bar{x}^2} - \bar{x}^2 \bar{Y} \simeq -\bar{x} \tag{38}$$

whose solution is

$$\bar{Y} = \frac{1}{2} \bar{x} \int_0^1 dt \, (1 - t^2)^{-1/4} \exp(-t\bar{x}^2 / 2). \tag{39}$$

We may also consider the approximate solution

$$\bar{Y} \simeq \frac{\alpha_1 \bar{x} + \bar{x}^2 / 6}{1 + \bar{x}^4 / 6}. \tag{40}$$

Finally, if we refer to Eq. (30) and combine it with Eq. (37) we obtain

$$\tilde{\xi}_x \simeq -\frac{\bar{Y}}{(k_y \delta_T)(dB_y/dx)} \frac{i\tilde{B}_x^0}{G_T} \tag{41}$$

where

$$G_T \equiv \left[\frac{d}{dx}(\Delta T)_e\right] / \left(\frac{dT_{e\|}}{dx}\right). \tag{42}$$

## 6. Total Momentum Conservation Equation

Referring to the total momentum conservation equation we take the $\mathbf{e}_z \cdot \nabla \times$ component of it [5] and obtain, within the $\delta_T-$ region,

$$-\omega_0(\omega_0 - \omega_{di})\rho \frac{d^2\tilde{\xi}_x}{dx^2} \simeq \frac{i}{4\pi}\left\{(\mathbf{k}\cdot\mathbf{B})\frac{d^2\tilde{B}_x}{dx^2} - k_y \frac{d^2 B_y}{dx^2}\tilde{B}_x\right\} \\ -\frac{1}{4\pi}\frac{d}{dx}\left[k_y B_y (\mathbf{k}\cdot\mathbf{B})\Delta\tilde{p}\right], \tag{43}$$

where $\rho = m_i n$, $\Delta\tilde{p} = 4\pi(\tilde{p}_{e\|} - \tilde{p}_{e\perp})/B^2$ and $\omega_{di} = \mathbf{k}\cdot\mathbf{u}_i$. Since $\beta_e = 8\pi n T_{e\|}/B^2 \ll 1$, we argue that the last term on the r.h.s. of Eq. (43) can be neglected. In view of the asymptotic matching condition discussed in the next section we limit consideration to the part of the solution within the $\delta_T-$ region for which $\tilde{\xi}_x(\bar{x})$ is odd in $\bar{x}$ and $\tilde{B}_x$ is even. Thus, for the following analysis,

$$-\omega_0(\omega_0 - \omega_{di})\rho \frac{d^2\tilde{\xi}_x}{dx^2} \simeq \frac{i}{4\pi}(\mathbf{k}\cdot\mathbf{B})\frac{d^2\tilde{B}_x}{dx^2} \tag{44}$$

is considered.

## 7. Asymptotic Matching $(T_{e\perp} \neq T_{e\|})$

If we refer to Eq. (44) and define

$$\omega_H^2 \equiv \frac{(B_y')^2}{4\pi\rho}, \quad \varepsilon_T \equiv \frac{\omega_0^2}{\omega_H^2(k_y \delta_T)^2} \tag{45}$$

and

$$\Lambda_0 \equiv \frac{dT_{e\|}/dx}{d(T_{e\|}-T_{e\perp})/dx}\left(1-\frac{\omega_{di}}{\omega_0}\right), \tag{46}$$

Eq. (44) can be rewritten as

$$\Lambda_0 \frac{d^2\bar{Y}}{d\bar{x}^2} \simeq \bar{x}\frac{d^2\varphi_0}{d\bar{x}^2}. \tag{47}$$

The asymptotic matching condition for the solution within the $\delta_T$ − region with that valid in the outer ideal MHD region, where $\tilde{B}_x = i(k_\| B)\tilde{\xi}_x$, and

$$0 \simeq (\mathbf{k}\cdot\mathbf{B})\left\{\frac{d^2\tilde{B}_x}{dx^2}-k^2\tilde{B}_x-\mathbf{k}\cdot\frac{d\mathbf{B}}{dx}\tilde{B}_x\right\}, \tag{48}$$

requires that

$$\Delta' = \frac{1}{\tilde{B}_x^0}\int_{\delta_T}dx\frac{d^2\tilde{B}_x}{dx^2} \simeq \frac{\Lambda_0\varepsilon_T}{\delta_T}\int_{-\infty}^{+\infty}d\bar{x}\frac{1}{\bar{x}}\frac{d^2\bar{Y}}{d\bar{x}^2} \equiv -D'_0\mathcal{J}_0, \tag{49}$$

where $\Delta' \equiv (1/\tilde{B}_x^0)\left[d\tilde{B}_x/dx\big|_{x=x_{0+}} - d\tilde{B}_x/dx\big|_{x=x_{0-}}\right]$, derived from the solution of Eq. (48),

$$D'_0 \equiv -\frac{\varepsilon_T}{\delta_T}\left(1-\frac{\omega_{di}}{\omega_0}\right)\frac{1}{G_T(x_0)} \tag{50}$$

and the integral $\mathcal{J}_0$, defined as $\mathcal{J}_0 \equiv -\int_{-\infty}^{+\infty}d\bar{x}\left(d^2\bar{Y}/d\bar{x}^2\right)/\bar{x} = \int_{-\infty}^{+\infty}d\bar{x}\left(1-\bar{x}\bar{Y}\right)$, when evaluated using Eq. (39), is $\mathcal{J}_0 = \sqrt{2}\left[\Gamma(3/4)\right]^2$.

Therefore, if $G_T(x_0)>0$, $D'_0<0$ and $\Delta'$ needs to be negative. This implies that the amplitude of $\tilde{B}_x$ has to be maximum within the reconnection region and, considering the terms entering the equation for $\tilde{B}_x$ in the outer region, corresponds to relatively large values of $k_y$.

If, instead $G_T(x_0)<0$ corresponding to a stronger decay (in $x$) of $T_{e\perp}$ than that of $T_{e\|}$, $\Delta'$ is required to be positive. Therefore, the role of $dJ_z/dx$ becomes important combined with the effects of the temperature anisotropy gradient. In fact, in this case a growth rate associated with the finite electrical resistivity can be found.

A case that deserves special attention is that of weak anisotropy that is $G_T = -\varepsilon_A$ with $\varepsilon_A < 1$. Then the required $\Delta'$ will have to be relatively large, corresponding to sharp current density gradients. On the other hand, if $G_T(x_0) > 0$, it becomes important to include the contribution of particle transport across the magnetic field in the derivation of Eq. (30) as shown in Appendix A. The possibility that this may introduce a growth rate depends on the form of the transport equation that can be adopted.

## 8. Effects of Finite Resistivity

Referring to the case where $G_T(x_0) < 0$, if the presence of a small resistivity is considered, $\omega$ will acquire an imaginary component that is $\omega = \omega_0 + i\gamma$ with $\gamma \ll \omega_0$. Then, while Eq. (20) remains valid to zeroth order in $\gamma/\omega_0$, the next order relevant equation is

$$\gamma \tilde{B}_x^0 \simeq i(\mathbf{k}\cdot\mathbf{B})\left\{-i\omega_0\tilde{\xi}_1 + \gamma\tilde{\xi}_0 + ik_y D_B \frac{1}{nT_{e\parallel}}(\tilde{p}_{e\parallel} - \tilde{p}_{e\perp})_1 + D_m \frac{d^2\tilde{B}_x}{dx^2}\right\}, \tag{51}$$

where $\mathrm{Im}(\tilde{\xi}_1/\tilde{B}_x^0) = 0$ and $D_m$ is the magnetic diffusion (resistive) coefficient. Then, we may argue that

$$\tilde{B}_x \simeq \tilde{B}_x^0\left[1 + \varepsilon_T \varphi_0(\overline{x})\right] + i\tilde{B}_1(\overline{x}) \tag{52}$$

where $|\tilde{B}_1/\tilde{B}_x^0| \sim \gamma/\omega_0 \ll \varepsilon_T$ and in this case

$$\gamma \simeq \frac{1}{\delta_T^2} D_m \varepsilon_T \left.\frac{d^2\varphi_0}{d\overline{x}^2}\right|_{\overline{x}=0} > 0. \tag{53}$$

We may consider $|(p_{e\perp})_1/(p_{e\parallel})_1| \ll 1$ and find

$$\tilde{\xi}_1 \simeq \frac{1}{\omega_0} \frac{\tilde{B}_x^0}{(\mathbf{k} \cdot \mathbf{B})} \left\{ \gamma - D_m \varepsilon_T \frac{1}{\delta_T^2} \frac{d^2 \varphi_0}{d\bar{x}^2} \right\} - i \frac{\gamma}{\omega_0} \tilde{\xi}_0 + \frac{k_y D_B}{\omega_0} \frac{1}{nT_{e\parallel}} (\tilde{p}_{e\parallel})_1 . \qquad (54)$$

Clearly, in order to evaluate $(p_{e\parallel})_1$ and continue the relevant analysis would require that the electron thermal energy balance equation and its solution be reconsidered taking Eq. (52) into account for the difference $\tilde{B}_x - \tilde{B}_x^0$. The total momentum conservation will have to be reconsidered accordingly.

## 9. Final Considerations

The analysis that has been carried out concerns the onset of a cyclic reconnection process that can take place in inhomogeneous plasmas when the electron distribution in momentum space departs from a Maxwellian. The simplest form of such a departure is an anisotropy of the electron temperature relative to the magnetic field. The theory includes the case where this anisotropy is small and involves a thermal energy balance equation for the longitudinal electron temperature featuring a larger longitudinal thermal conductivity than that transverse to the magnetic field. The relevant transport coefficients may in fact be associated with a pre-existent state of "turbulence" produced by microscopic modes.

Referring to collisionless regimes, a parallel "laminar" analysis needs to be carried out, for a comparison, involving the direct effects of resonant mode-particle interactions for the unperturbed configuration and conditions described in Section 2 following that presented in Ref. [7]. We observe that in view of the results presented in the previous sections, the analysis given in Ref. [3] can be considered valid for $T_{e\perp} \ll T_{e\perp} \simeq T_e$.

In a different context, a "primary" (slow) cosmic alternator has been considered as a process to create nearly stationary sheared magnetic field configurations that can undergo reconnection and become amplified in the presence of an electron temperature anisotropy. In particular, the primary alternator is proposed to be sustained by inhomogeneous density and electron temperature fluctuations emerging from a circumbinary disk [8] associated with a binary system (e.g. involving black holes).

**Acknowledgements**

This work was supported in part by the Kavli Foundation (through M. I. T.) and by the Ministry for University and Research of Italy. It is a pleasure to thank Francesco Pegoraro and Valeria Ricci for their active interest in the subject of this paper.

**Appendix**

**Particle Transport**

For the particle transport flux in the $x-$ direction, denoted by $\Gamma_p$, we adopt the transport equation introduced in Ref. [6] that has been found to be consistent with the plasma density profiles produced by numerous experiments. That is

$$\Gamma_p \simeq -D_p \frac{dn}{dx} - V_p n, \tag{A-1}$$

where $V_p$ is the inward transport velocity counteracting diffusion. Consequently, for the stationary state

$$\frac{d\Gamma_p}{dx} \simeq -D_p \frac{d^2 n}{dx^2} - \left(V_p + \frac{dD_p}{dx}\right)\frac{dn}{dx} - \frac{dV_p}{dx} n = S_p, \tag{A-2}$$

where $n$ is assumed to be a monotonically decreasing, even function of $x$, $S_p$ is the particle source considered, like the relevant diffusion coefficient $D_p$, as an even function of $x$ while $V_p$ is odd [6]. Then the perturbed form of $d\Gamma_p/dx$ that can be adopted is

$$\frac{\partial \hat{\Gamma}_p}{\partial x} \simeq -D_p \frac{\partial^2 \hat{n}}{\partial x^2} - \left(V_p + \frac{dD_p}{dx}\right)\frac{\partial \hat{n}}{\partial x} - \frac{dV_p}{dx} \hat{n} \tag{A-3}$$

with the mass conservation equation

$$-i\bar{\omega}_e \hat{n} + \hat{u}_{ex} \frac{dn}{dx} + \frac{\partial \hat{\Gamma}_p}{\partial x} \simeq 0. \tag{A-4}$$

If we assume that $D_p / |\bar{\omega}_e \delta_T^2| \ll 1$, to the lowest order in this small parameter, we have $\tilde{n} \simeq -(dn/dx)\tilde{\xi}_0$ where $\tilde{\xi}_0$ is an odd function of $(x-x_0)$ over the $\delta_T$-region. Then, following Eq. (24),

$$\tilde{u}_{ex} \simeq -i \frac{k_y D_B}{T_{e\parallel}} \left\{ (\Delta T)_e \left(-\frac{i}{\bar{\omega}_e n}\right) \left[\tilde{u}_{ex} \frac{dn}{dx} + \frac{d\tilde{\Gamma}_p}{dx}\right] + \tilde{T}_{e\parallel} \right\} \tag{A-5}$$

and, instead of Eq. (29),

$$\frac{d}{dx}\left[n(\Delta T)_e \tilde{\xi}_x\right] \simeq (\Delta T)_e \left(\frac{dn}{dx} \tilde{\xi}_x\right) - \tilde{T}_{e\parallel} + (\Delta T)_e \left(\frac{i}{\bar{\omega}_e} \frac{d\tilde{\Gamma}_p}{dx}\right) \tag{A-6}$$

that gives

$$\left[\frac{d}{dx}(\Delta T)_e\right]\tilde{\xi}_x \simeq -\tilde{T}_{e\|} + i(\Delta T)_e \frac{1}{\bar{\omega}_e n}\frac{d}{dx}\tilde{\Gamma}_p. \tag{A-7}$$

Further analysis indicates that the contribution of $d\tilde{\Gamma}_p/dx$ is significant in order to find $\operatorname{Im}\omega$ in the case where $G_T(x_0) > 0$, but the sign of this depends on the multiplicity of factors entering Eq. (A-2) and cannot be considered conclusive.